\documentclass[aps,pra,reprint,groupedaddress,floatfix]{revtex4-1}

\usepackage[T1]{fontenc}
\usepackage{amsmath}
\usepackage{amssymb}
\usepackage{graphicx}
\usepackage{braket}
\usepackage{color}

\newcommand{\adag}{\ensuremath{\hat{a}^{\dagger}}}
\newcommand{\bdag}{\ensuremath{\hat{b}^{\dagger}}}
\newcommand{\rmd}{\ensuremath{\mathrm{d}}}
\newcommand{\rmi}{\ensuremath{\mathrm{i}}}

\newcommand{\op}{\ensuremath{\omega_\mathrm{p}}}
\newcommand{\os}{\ensuremath{\omega_\mathrm{s}}}
\newcommand{\oi}{\ensuremath{\omega_\mathrm{i}}}

\newcommand{\opz}{\ensuremath{\omega_\mathrm{p}^{(0)}}}
\newcommand{\osz}{\ensuremath{\omega_\mathrm{s}^{(0)}}}
\newcommand{\oiz}{\ensuremath{\omega_\mathrm{i}^{(0)}}}

\newcommand{\nus}{\ensuremath{\nu_\mathrm{s}}}
\newcommand{\nui}{\ensuremath{\nu_\mathrm{i}}}
\newcommand{\kp}{\ensuremath{k_\mathrm{p}}}
\newcommand{\ks}{\ensuremath{k_\mathrm{s}}}
\newcommand{\ki}{\ensuremath{k_\mathrm{i}}}
\newcommand{\dkp}{\ensuremath{k_\mathrm{p}^{(1)}}}
\newcommand{\dks}{\ensuremath{k_\mathrm{s}^{(1)}}}
\newcommand{\dki}{\ensuremath{k_\mathrm{i}^{(1)}}}

\newcommand{\nps}{\ensuremath{n_\mathrm{ps}}}
\newcommand{\npi}{\ensuremath{n_\mathrm{pi}}}
\newcommand{\nsi}{\ensuremath{n_\mathrm{si}}}

\newcommand{\taus}{\ensuremath{\tau_\mathrm{s}}}
\newcommand{\taui}{\ensuremath{\tau_\mathrm{i}}}
\newcommand{\wigner}{\ensuremath{W_\mathrm{PDC}(\os,\oi,\taus,\taui)}}
\newcommand{\nuwigner}{\ensuremath{W_\mathrm{PDC}(\nus,\nui,\taus,\taui)}}

\newcommand{\wignersc}{\ensuremath{W_\mathrm{s}^{\mathrm{(cond)}}(\nus,\taus)}}
\newcommand{\wignersu}{\ensuremath{W_\mathrm{s}^{\mathrm{(uncond)}}(\nus,\taus)}}

\newcommand{\rhopdc}{\ensuremath{\hat{\rho}}}

\begin{document}

%Title of paper
\title{Characterizing entanglement in pulsed parametric downconversion using chronocyclic Wigner functions}

\author{Benjamin Brecht, and Christine Silberhorn}
\affiliation{Integrated Quantum Optics, University of Paderborn, Germany}

\date{\today}

\begin{abstract}
	Pulsed parametric downconversion (PDC) processes generate photon pairs with
	a rich spectral-temporal structure, which offer an attractive
	potential for quantum information and communication applications. In this
	paper, we investigate the four-dimensional chronocyclic Wigner function
	$\wigner$ of the PDC state, which naturally
	lends itself to the pulsed characteristics of these states. From this
	function we derive the conditioned time-bandwidth
	product of one of the pair photons, a quantity which is not only a valid
	measure of entanglement between the PDC photons but also allows to
	highlight a remarkable link between the discrete and continuous
	variable descriptions of PDC. We numerically analyze PDC processes with
	different conditions to	demonstrate the versatility of our approach, which
	is applicable to a large number of current PDC sources.
\end{abstract}

% insert suggested PACS numbers in braces on next line
\pacs{}
\maketitle
\section{Introduction}
In today's quantum optical applications, parametric downconversion (PDC) 
sources are a well-established tool for the generation of a large variety
of quantum states. Their versatility covers the heralding
of single photons, the creation of highly entangled
photon-pair states and the generation of bright single-mode as well as two-mode 
squeezed beams of light \cite{Kwiat:1995ck,Slusher:1987in,Anderson:1995cg,Zhang:2000fi}. 
Pumping the PDC sources with
spectrally broad ultrafast pump pulses further increases the repertoire of
realizable quantum states, including the generation of genuine
single-photon quantum pulses \cite{Mosley:2008ir} or the creation of multimode
quantum frequency combs \cite{deValcarcel:2006uh,Pinel:2012ca}.

Naturally, the experimental progress during recent
years has been accompanied by elaborate
theoretical considerations which aim for a complete understanding of the PDC
process \cite{Caves:1991wd,Grice:1997tk,URen:2005wb,Wasilewski:2006ht,
Mikhailova:2008dz,Mosley:2009kr}. 
However, when it comes to the description of PDC
output states, two at a first glance disparate methods are still prevailing.

In continuous variable quantum optics, quantum states are typically described
by their Wigner function and the analysis concentrates on
evaluating the fluctuations of two conjugate phase-space quadratures. 
Non-classical features are mostly associated with negative values of the
Wigner function or with quadrature fluctuations smaller than those
of a coherent state. In this respect, non-degenerate PDC states exhibit reduced
joint fluctuations of the conjugate $(X_\mathrm{s} + X_\mathrm{i})$ and 
$(Y_\mathrm{s} - Y_\mathrm{i})$ quadratures, and can thus overcome an apparent
Heisenberg's uncertainty relation. This feature is known as two-mode
squeezing and is an intuitive way for depicting the entanglement which is
generated in a PDC process.

In contrast, in the context of discrete variable systems research mainly focusses
on the photon-pair picture, neglecting higher order photon-number 
contributions but taking into account modal characteristics. The spectral-temporal properties of PDC states are commonly described by their
complex-valued joint spectral amplitude (JSA) function $f(\os,\oi)$, from which also the entanglement between signal and idler can be retrieved.
However, this description has one major drawback. 
Generally the measurements used to determine the joint spectral distribution 
of a PDC state are phase-insensitive intensity measurements and thus do not yield the JSA, but the joint spectral intensity (JSI) function. Therefore, any information hidden in the phase term of the JSA gets lost during measurement. This can be deceptive when trying to generate decorrelated PDC states, which are highly valuable for the heralded generation of pure single photons 
\cite{Mosley:2008ir, Shi:2008tl,Eckstein:2011wl}.
By judging the purity of the heralded photon from spectral intensity
measurements only, one possibly overestimates the performance of the heralded
single-photon source.

In this paper we combine the advantages of both, continuous variable and
discrete variable systems, into another approach towards describing the spectral-temporal behaviour of PDC states. We utilize the chronocyclic Wigner function formalism which is well-known in classical ultrafast optics, where it is routinely used to describe the spectral-temporal properties of pulses \cite{Paye:1992ue} and forms the basis of ultrafast pulse-characterization schemes \cite{Iaconis:1998vr}. Here we apply it to a PDC pumped by an ultrafast pulse. Note that this approach naturally lends itself to the pulsed characteristics of PDC sources and -- since the Wigner function is real-valued -- all quantities are accessible by direct measurements of the respective time and frequency distributions. We present a compact analytic
expression for the Wigner function which is valid for a large number of current
PDC sources and introduce the concept of a conditioned time-bandwidth 
product (TBP). In classical optics, the TBP of a pulse is ultimately bounded from below by the Fourier limit. However, this paradigm does not hold true in the quantum regime. In particular we show that, for PDC states, spectral-temporal entanglement between signal and idler overcomes this classical boundary and the conditioned TBP forms a valid measure of entanglement. 
\section{Derivation of the chronocyclic Wigner function}
In order to derive an analytic expression for the chronocyclic PDC
Wigner function $\wigner$, we start by assuming a PDC state $\ket{\psi}$,
of the form

\begin{equation}
	\ket{\psi} = B\int\rmd\os\rmd\oi f(\os,\oi)\adag(\os)\bdag(\oi)\ket{00}.
\end{equation}

\noindent Here, the parameter $B$ is an overall coupling constant, $\adag$ and $\bdag$ are the standard signal
and idler creation operators and the function $f(\os,\oi)$ is the complex-valued JSA
which fully characterizes the generated state \cite{Grice:1997tk}. Note that we neglect any higher
order photon number contributions of the PDC state which is a good
approximation in the limit of low pump powers, and that we restrict our
analysis to one dimension in space which is applicable in the case of
waveguided PDC.

The chronocyclic Wigner function can then be retrieved from the PDC density operator
$\rhopdc=\ket{\psi}\bra{\psi}$ by a two-dimensional Wigner transform

\begin{equation}
	\begin{split}
		W(&\os,\oi,\taus,\taui) = \frac{1}{(2\pi)^2}
		\int_{-\infty}^{\infty}\limits\mathrm{d}\omega'\mathrm{d}\omega''
		e^{\rmi\omega'\taus+\rmi\omega''\taui}\times\\
		&\times\left<\os-\frac{\omega'}{2},\oi-\frac{\omega''}{2}\right|
		\rhopdc\left|\os+\frac{\omega'}{2},\oi+\frac{\omega''}{2}\right>.
	\end{split}
	\label{eq:wigner_transform}
\end{equation}

Since we aim for presenting a compact analytical expression for $\wigner$,
we introduce two simplifications,
which do not limit our general theory. All calculations can be
performed numerically for cases in which our simplified model does not yield a satisfying description.

Firstly, we express the JSA in terms of Gaussian functions, with the 
constituents $\alpha(\os, \oi)$ called the pump envelope function, and 
$\phi(\os, \oi)$ called the phasematching function, which reflect energy
and momentum conservation of the PDC process, respectively.

\begin{equation}
	\begin{split}
		f(\os,\oi)&=\alpha(\os,\oi)\,\phi(\os,\oi)=\\
		&=\exp\left(-\frac{\Delta\omega^2}{2\sigma^2}-
		\rmi a\Delta\omega^2\right)\times\\
		&\times\exp\left[-\gamma\left(\frac{L}{2}
		\Delta k\right)^2\right]\exp\left(\rmi\frac{L}{2}\Delta k\right).
	\end{split}
\end{equation}
Here we introduced the abbreviation $\Delta\omega=\opz-\os-\oi$, which
denotes the difference between the central pump frequency $\opz$ and signal 
and idler frequencies $\os$ and $\oi$. The spectral width of the pump pulse is given by $\sigma$, the length of the waveguide by $L$ and $\Lambda$ denotes the periodic poling period, deployed to remove the phasemismatch $\Delta k=\kp(\op) - \ks(\os) - \ki(\oi) - \frac{2\pi}{\Lambda}$ between pump, signal and idler. In contrast to the the standard description of PDC, we explicitly take into account a temporal chirp of the pump pulse, characterized by the parameter $a$.

The approximation of the phasematching
with a Gaussian can experimentally be achieved by applying an appropriate spatial chirp to the poling 
period $\Lambda$ \cite{Branczyk:2011ti}. However, this simplification is a good approximation for PDC sources in general.

Secondly, we use a Taylor series expansion of the phasemismatch $\Delta k$
up to first order around the perfectly phasematched central frequencies $\opz$, 
$\osz$ and $\oiz$ and end up with \cite{Grice:9999ud}

\begin{equation}
	\Delta k\approx\left(\dkp-\dks\right)\nus+
	\left(\dkp-\dki\right)\nui,
	\label{eq:deltak}
\end{equation}

\noindent where $k^{(1)}_\mathrm{p, s, i}$ are the inverse group velocities of pump, signal and idler, given by the ratio between group refractive indices $n^{(g)}_\mathrm{p,s,i}$ and the speed of light. Note that we neglect second order contributions here, since the group-velocity dispersion of the crystal does not play a role for the PDC investigated here \cite{Brida:2009gb}.
In equation (\ref{eq:deltak}), the variables
$\nus=\osz-\os$ and $\nui=\oiz-\oi$ describe the frequency offsets of
signal and idler from their perfectly phasematched central frequencies.
By rewriting the JSA as a function of the frequency offsets $\nus$ and
$\nui$ we find

\begin{equation}
	\begin{split}
		f&(\nus,\nui) =\\ &=\exp\left[-\frac{(\nus+\nui)^2}{2\sigma^2}-
		\frac{\gamma L^2}{4c^2}\left(\nps\nus+\npi\nui\right)^2\right]\times\\
		&\times\exp\left[-\rmi a (\nus+\nui)^2+\rmi\frac{L}{2c}
		(\nps\nus+\npi\nui)\right],
	\end{split}
	\label{eq:jsa_ana}
\end{equation}

\noindent where we abbreviated $n_{ij} = n^{(g)}_i - n^{(g)}_j$ for
$i,j = \mathrm{p}, \mathrm{s}, \mathrm{i}$, and used
$k(\omega)=\frac{n(\omega)\omega}{c}$ with $c$ denoting the speed-of-light. 
After straightforward calculations employing equation
(\ref{eq:wigner_transform}) we derive a compact analytic expression for  the
four-dimensional chronocyclic PDC Wigner function:

\begin{widetext}
	\begin{equation}
		\begin{split}
			W(\nus,\nui,\taus,\taui)=&\sqrt{\frac{2}{\gamma}}\,
			\frac{|B|^2c\sigma}{L\pi|\nsi|}\,e^{-1/2\gamma}
			\times\exp\left[-\frac{1}{\sigma^2}
			(\nus+\nui)^2-\frac{\gamma L^2}{2c^2}(\nps\nus+\npi\nui)^2-
			4a^2\sigma^2(\nus+\nui)^2\right]\times\\
			&\times\exp\left[-\frac{2c^2}{\gamma L^2\nsi^2}(\taus-\taui)^2-
			\frac{\sigma^2}{\nsi^2}(\npi\taus-\nps\taui)^2+\frac{2c}
			{\gamma L\nsi}(\taus-\taui)\right]\times\\
			&\times\exp\left[\frac{4a\sigma^2}{\nsi}(\nus+\nui)
			(\npi\taus-\nps\taui)\right].
		\end{split}
		\label{eq:wigner_ana}
	\end{equation}
\end{widetext}

Note that a non-vanishing chirp of the pump pulse introduces spectral-temporal
correlations between the two sibling photons generated in the PDC which cannot
be observed by common measurements of the spectral intensity distribution.

From equation (\ref{eq:wigner_ana}) we can directly
deduce the limits of our second simplification. As soon as signal and idler
photons have similar group-velocities, $\nsi$ becomes very small and equation
(\ref{eq:deltak}) is not valid anymore, because higher order terms have to be
taken into account in the Taylor series expansion of the wavevectors.
Thus, the analytic expression is not valid
for degenerate type I PDC sources, where signal and idler share the same 
polarization. Numerical calculations have to be applied then. For most other
PDC sources based upon type II and non-degenerate type I processes, however, this analytic expression is valid and provides a practical approach for studying the bi-photon state.
\section{Entanglement between the PDC photons}
Having derived an analytic expression for the chronocyclic PDC Wigner
function, we now deploy it to analyze the entanglement between the signal and idler
photons created during the PDC process. We start by calculating
single-photon Wigner functions (SPWF) from the PDC Wigner function. Note that
detailed considerations on the SPWF have also been presented
in \cite{SanchezLozano:2011ta} investigating the single-photon purity of 
a heralded PDC photon under several experimental conditions.

Here, we concentrate on the striking similarity between the
spectral-temporal description of PDC states, common in discrete variable 
quantum optics and the Wigner formulation used in the field of
continuous variable quantum optics, by introducing the notion of a 
conditioned SPWF. To highlight the close relationship between the two 
approaches we first recall two-mode squeezing and the cooperativity, two
concepts associated with continuous and discrete variable approaches,
respectively. Thereafter we introduce the notion of the conditioned TBP
and point out its link to both representations.
\subsection{Continuous variable description}
In the context of continuous variable systems, PDC states are mostly understood
by deploying a four-dimensional Wigner function $W(X_1,Y_1,X_2,Y_2)$, where
$X_{1,2}$ and $Y_{1,2}$ are conjugate quadratures of signal and idler, 
respectively. The amount of two-mode squeezing $\zeta$ -- which is 	tantamount
with the amount of entanglement between the PDC photons -- can be determined 
when regarding joint fluctuations of signal and idler, as
\begin{equation}
	\Delta^2(X_1+X_2) =	\Delta^2(Y_1-Y_2) = \exp(-2|\zeta|)\leq1.
\end{equation}
Thus, when fixing $X_2$ and $Y_2$ to distinct values $X_2^{(0)}$ and 
$Y_2^{(0)}$, $\Delta^2(X_1+X_2^{(0)})\Delta^2(Y_1-Y_2^{(0)})$ can overcome
Heisenberg's uncertainty limit. The fact that two quadratures of the same
field of a PDC state can apparently be defined with arbitrary precision --
or at least below the Heisenberg limit \cite{Reid:1988js} --
exemplifies the EPR paradox \cite{Einstein:1935tq} and
has first been demonstrated in \cite{OuZY:1992wj}.
\subsection{Discrete variable description}
The typical representation of the same state becomes quite different, when
employing the discrete variable description of PDC for bi-photon states. 
In contrast to the previous definition, higher photon number
contributions are normally neglected in this approach.
Extracting information on the
entanglement between signal and idler photons is typically accomplished by means of the Schmidt decomposition
of the JSA function \cite{Law:2000wd}, where $f(\os,\oi)$ is decomposed
into two sets of orthonormal basis functions, such that
\begin{equation}
	f(\os,\oi) = \sum_n\lambda_n\phi(\os)\theta(\oi)
\end{equation} 
and the Schmidt coefficients $\lambda_n$ fulfill the condition 
$\sum_n\lambda_n^2=1$. This, in term, allows us to determine the so-called cooperativity 
$K=\sum_n 1/\lambda_n^4$, a quantity representing an established measure of entanglement
between the PDC photons. When the photons are uncorrelated, the JSA function
is separable and only one basis mode for signal and idler remains. Consequently,
the cooperativity then equals one. With increasing entanglement between signal
and idler, the cooperativity increases and approaches infinity for perfectly correlated photon pairs.
\subsection{Four-dimensional chronocyclic Wigner function}
We can now find an intuitive connection between both, two-mode squeezing
and the cooperativity. To this end we consider on the one hand the unconditioned chronocyclic SPWF, which we obtain by ignoring any knowledge about one of the two photons. On the other hand, we calculate the conditioned chronocyclic SPWF by fixing the arrival time and frequency offset of one photon. The two functions are then given by:
\begin{align}
	&\wignersu = \int\rmd\nui\rmd\taui\nuwigner,\label{eq:uspwf}\\
	&\wignersc = W_\mathrm{PDC}(\nus, \taus; \nui=\nui^{(0)}, 
	\taui=\taui^{(0)}).
	\label{eq:cspwf}
\end{align}

In any of the two cases, the SPWF is described by a two-dimensional Gaussian
function in the $(\nus,\taus)$-plane. Retrieving the TBP
$\Delta\nus\Delta\taus$ from this function is a matter of simple
geometric considerations, which are detailed in \cite{Laiho:2009vh}. Here we focus on its significance for new insights on the underlying physics of the generated state.

Because time and frequency share a Fourier relationship, a given spectral width
of a light pulse enforces a minimum duration due to $\Delta\nu\Delta\tau
\geq const$, where the value of the constant depends on the pulse shape and
the employed width-parameter of the pulse. We chose our
normalization such that the Fourier relationship can conveniently be written
as $\Delta\nu\Delta\tau\geq1$. The TBP measures the
similarity of the light pulse to its idealized version and is thus a measure
of pulse quality. It has been shown in \cite{SanchezLozano:2011ta} that
for single-photon pulses a TBP which equals the lower bound is tantamount to a
pure single-photon, whereas a larger TBP suggests impurity.

However, when employing entangled photon-pair states, we can also analyze the
conditioned TBP of one of the pair photons. This expression can be retrieved
from the conditioned SPWF presented in equation (\ref{eq:cspwf}). By
doing so a classically surprising and counter-intuitive property emerges:
the conditioned TBP can actually become smaller than one, which is forbidden
for classical light pulses. A similar phenomenon
occurs in the case of two-mode squeezed states, where the joint fluctuations
can overcome an apparent Heisenberg uncertainty relation, which is a
fingerprint of the quantum feature of entanglement between signal and idler.

For the upcoming analysis, we write down the analytical expression for the
inverse conditioned TBP (ICTBP) that is $(\Delta\nu\Delta\tau)^{-1}$:
\begin{equation}
	\mathrm{ICTBP} = \sqrt{\frac{\npi^2+\nps^2}{\nsi^2}+\frac{\gamma
	L^2\sigma^2\npi^2\nps^2}{2c^2\nsi^2}+\frac{2c^2(1+4a^2\sigma^4)}
	{\gamma L^2\sigma^2\nsi^2}}.
\end{equation}
If the PDC photons are uncorrelated, the ICTBP equals one. For increasing
correlations between signal and idler, the violation of the Fourier-relationship
becomes stronger and the ICTBP increases.

\section{Analysis of the chronocyclic Wigner function}
In this section, we evaluate the chronocyclic PDC Wigner function $\wigner$
for two distinct cases of PDC processes. First, we concentrate on the case
of spectrally decorrelated PDC which has received a lot of attention in recent
years, as it allows for the direct heralding of pure single photon pulses
without the need for any filtering \cite{Mosley:2008ir, Shi:2008tl,
Eckstein:2011wl}. However, the implementation of these kinds of PDC sources
still is a major experimental challenge and in general PDC processes exhibit
strong spectral correlations between signal and idler. Hence, we investigate,
as a second case, a correlated PDC process. 

In the following we assume the PDC
processes to be pumped by Fourier limited pump pulses, which do not comprise
any temporal chirp. Moreover, as a testbed, we choose the PDC process first
presented in  \cite{Eckstein:2011wl}. It is realized in a KTP waveguide 
pumped with ultrafast pump pulses around $775\,$nm and can generate
decorrelated photons in orthogonal polarizations, centered around $1550\,$nm.
By changing the spectral width of the pump pulses, the spectral-temporal 
correlations between signal and idler photons can smoothly be tuned. 

\begin{figure}
	\includegraphics[width=\linewidth]{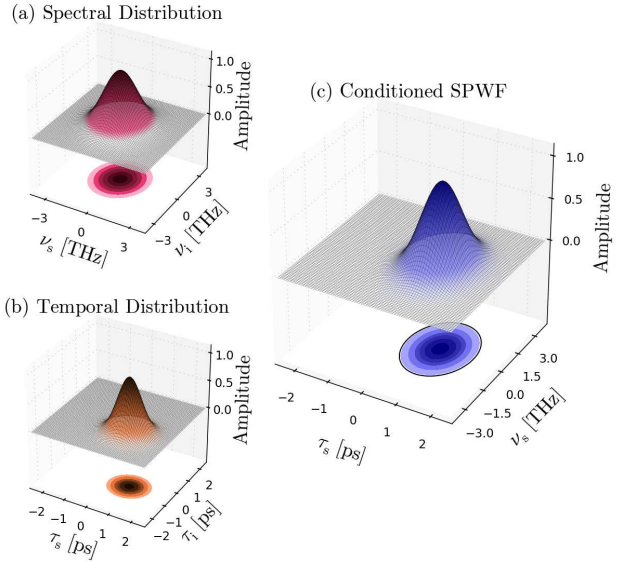}
	\caption{(a) JSI of the decorrelated PDC process. It is not possible to
	infer the signal frequency from measuring the idler frequency. 
	(b) Corresponding JTI, exhibiting no correlations between signal
	and idler photon arrival times. (c)	Conditioned SPWF of the signal photon.
	For further information see text.}
	\label{fig:case_1}
\end{figure}

\subsection{Spectrally decorrelated PDC}

\noindent In figures \ref{fig:case_1}(a) and \ref{fig:case_1}(b), we plot the JSI and the JTI of the spectrally decorrelated PDC process, respectively. Obviously, no information on the signal can be gained from measuring the idler frequency offset or arrival time, meaning that signal and idler pulses are generated in Fourier limited pulses with flat phase distribution.
Note that the offset of the JTI from the center of the figure reflects the different group velocities of signal and idler in the nonlinear waveguide.

Figure \ref{fig:case_1}(c) shows the conditioned SPWF, where we fixed the
idler frequency offset and arrival time to zero. This choice is arbitrary
and does not influence the shape of the conditioned SPWF, as long as the values
are well inside the idler spectrum and duration. The black circle indicates the
$1/e^2-$width of the unconditioned SPWF. Obviously, the
conditioned and unconditioned SPWF are similar. In this case, the ICTBP equals
one,  which corresponds to the signal photon residing in a Fourier limited pulse.
This result has been presented in \cite{SanchezLozano:2011ta}, and it serves here as a cross-check for our approach.

\begin{figure}
	\includegraphics[width=\linewidth]{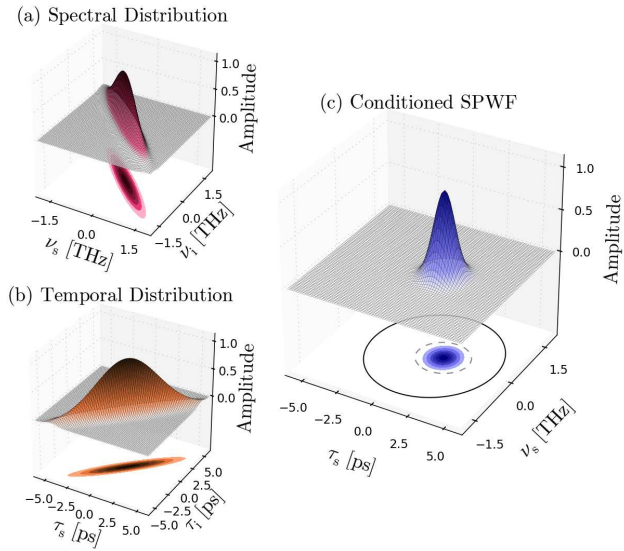}
	\caption{(a) JSI of the correlated PDC process. (b) Corresponding JTI,
	exhibiting inverted correlations. (c) Conditioned SPWF of
	the	signal. For further information see text.}
	\label{fig:case_2}
\end{figure}

\subsection{Spectrally correlated PDC}

\noindent We now turn our attention to the more interesting and common case of a
spectrally correlated PDC process, where we can infer the signal frequency
from a measurement of the idler frequency.

In figures \ref{fig:case_2}(a) and \ref{fig:case_2}(b), we show the JSI and
JTI of the spectrally correlated PDC, respectively. It can be nicely seen that
spectral anti-correlations correspond to temporal correlations, as expected for the two-dimensional Fourier transform between the two domains.

In figure \ref{fig:case_2}(c) we plot the conditioned SPWF of the
signal, where the black solid line indicates again the $1/e^2-$width of the
unconditioned SPWF. In addition, the grey dashed line indicates the 
$1/e^2-$width of the SPWF of a pure single photon, which exhibits a TBP of one.
We can qualitatively deduce the TBP from the size of the SPWF and find that
the TBP of the unconditioned signal is obviously larger than one. Note that this corresponds to a mixed quantum state. In contrast, the TBP of the conditioned signal is smaller than one, which is an indicator for the entanglement between signal and idler.

\section{Analysis of the conditioned TBP}

\noindent In this section we deploy again the PDC process from \cite{Eckstein:2011wl}
to actually calculate the ICTBP, as well as the cooperativity $K$. For the latter, we
employ two approaches: on the one hand we calculate $K$ directly from the
JSA given in equation (\ref{eq:jsa_ana}), on the other hand we calculate
$K$ from the JSI, which mimics the usual situation in the laboratory.
The JSI is found by
\begin{equation}
	F(\nus,\nui) = \left|f(\nus,\nui)\right|^2,
\end{equation}
where we mention again that any phase information gets lost during this step. We
concentrate on two scenarios. In the first case, we consider Fourier limited
pump pulses, whereas in the second case we turn our attention towards the more
realistic case of pump pulses, which exhibit a temporal chirp. To simulate
different degrees of correlation between signal and idler, we change the
spectral width of the pump pulses $\sigma$. Note that the JSI becomes
decorrelated for a pump pulse FWHM of around $2\,$nm.

\begin{figure}
	\includegraphics[width=\linewidth]{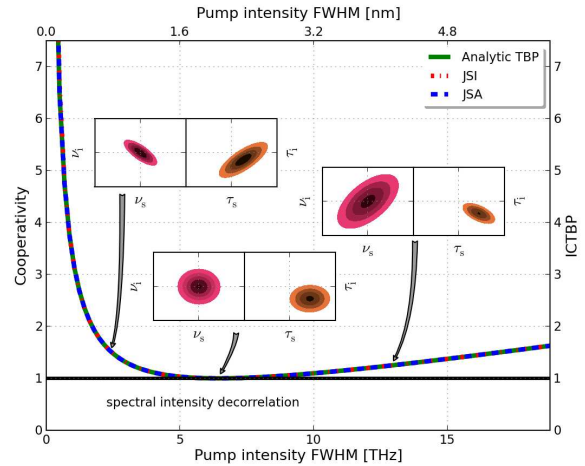}
	\caption{The ICTBP and the cooperativity $K$, calculated from
	the JSA and the JSI function for different degrees of spectral-temporal
	correlation between signal and idler. The insets show the respective JSI
	(left) and JTI (right) for three distinct spectral widths of the pump
	pulses.	Further information see text.}
	\label{fig:ctbp_1}
\end{figure}

In figure \ref{fig:ctbp_1}, we plot the ICTBP and the cooperativity $K$ for
the case of Fourier limited pump pulses. The insets show the JSI (left) and JTI
(right) for three different spectral widths of the pump. As expected, the cooperativities calculated from JSA and JSI are equal, since in this situation no spectral-temporal correlations are encoded on the phase of the state. The ICTBP exactly equals the cooperativities, which justifies its proposed role as measure of entanglement.

In the following, we investigate, if the similarity between ICTBP and $K$
persists when introducing a chirped pump and thus spectral-temporal correlations to the state. To this end, 
we consider a pump chirp of $3\cdot10^5\mathrm{fs}^2$ to clearly visualize its impact. Lower values of pump chirp decrease the investigated effects, but do not completely suppress them. 

\begin{figure}
	\includegraphics[width=\linewidth]{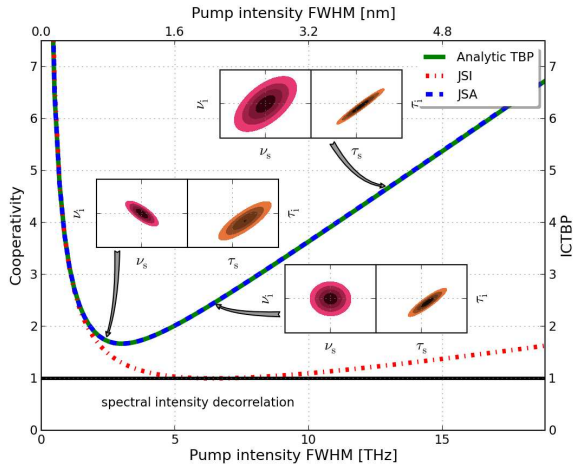}
	\caption{The ICTBP and the cooperativity $K$, calculated from
	the JSA and the JSI function for different degrees of spectral-temporal
	correlation between signal and idler under the assumption of a chirped pump
	pulse. The insets show the respective JSI (left)
	and JTI (right) for three distinct spectral widths of the pump pulses.
	Further information see text.}
	\label{fig:ctbp_2}
\end{figure}

In figure \ref{fig:ctbp_2} we plot again the ICTBP and the cooperativity $K$.
The insets show the JSI (left) and JTI (right) for the same spectral pump widths as in figure \ref{fig:ctbp_1}.
Because of the spectral-temporal correlations, we now find strongly correlated signal and idler arrival times even for a decorrelated JSI. For the considered pump chirp, this holds also true for cases where we would usually expect an anti-correlated JTI, as depicted in the upmost inset. Since the pump chirp enters equation (\ref{eq:jsa_ana}) in a quadratic phase term, we do not see its effect in intensity measurements. Therefore, the cooperativity calculated from the JSI strongly deviates from the one retrieved from the JSA. In contrast, the ICTBP again equals the phase-sensitive cooperativity from the JSA. Thus, full information on the entanglement between signal and idler can be gained from the PDC Wigner function $\wigner$. As a practical note we point out that this knowledge can in principle be obtained from spectral and temporal intensity measurements, but only if both, JSI and JTI, are measured.

Finally, we want to draw attention towards the minimum in the ICTBP. In the
case of Fourier limited pump pulses depicted in figure \ref{fig:ctbp_1},
the minimum value of ICTBP=1 is reached for spectral and temporal
decorrelation of signal and idler. However, as soon as the pump exhibits
a temporal chirp, the position of the minimum moves towards stronger spectral
anti-correlations, which partially compensate for the temporal correlations
introduced by the chirp. In figure \ref{fig:ctbp_2}, the minimum of the
ICTBP is at the point, where the spectral anti-correlations are about as
strong as the temporal correlations, as depicted in the leftmost inset. 
\section{Conclusion}
In this paper we have brought together well-known concepts from the
discrete and the continuous variable description of PDC to form an intuitive
and complete description of the resulting state. We have derived
a compact analytic expression for the four-dimensional
chronocyclic Wigner function $\wigner$ of a PDC state, where we included
the effects of different group-velocities of pump, signal and idler fields
and the effects of chirped pump pulses. In particular for the case of a
pulsed pump, this description naturally lends itself to the ultrafast
characteristics of the generated signal and idler.

Utilizing this expression, we have introduced the ICTBP of one of the generated
PDC fields. We have shown that this quantity exactly equals the cooperativity
$K$, which can be obtained from the JSA of the PDC state and thus forms a valid
measure of entanglement between signal and idler. Moreover, we have shown that,
given entanglement between signal and idler, the conditioned TBP becomes smaller
than the classical Fourier limit. This surprising feature is similar to the phenomenon
of two-mode squeezing in the continuous variable description of PDC, where the
conditioned quadrature fluctuations overcome an apparent Heisenberg's
uncertainty limit and highlights the similarity between the two seemingly 
disparate descriptions of PDC.

We have analyzed the ICTBP for different degrees of correlation between signal
and idler, and for situations with and without pump chirp, respectively.
From the results, we could show that it is not sufficient to only measure the
JSI or JTI of a PDC state to characterize the entanglement between signal
and idler. One either has to measure both degrees of freedom, or perform
experimentally highly challenging, phase-sensitive measurements in time or
frequency.
\section*{Acknowledgements}
The authors thank Andreas Christ and Georg Harder for fruitful discussions
and helpful comments. 
The research leading to these results has received funding from the 
European Community's Seventh Framework Programme FP7/2001-2013 under 
grant agreement no. 248095 through the Integrated Project Q-ESSENCE.
\bibliography{/Users/bbrecht/work/tex_stuff/bibliography}
\end{document}